\documentclass[prb,showpacs]{revtex4}
\pdfoutput=1

\usepackage{graphicx}

\begin{document}

\title{Climate Sensitivity and the Response of Temperature to CO$_2$}

\author{Arthur P. Smith}
\affiliation{apsmith@altenergyaction.org}

\begin{abstract}
A review of some of the evidence for the IPCC's conclusion
that doubling CO$_2$ levels will warm Earth significantly,
in contrast to the claims of a recent article\cite{monckton}.
Simply looking at raw temperature and CO$_2$ data over the
past 150 years gives a transient response of roughly 2 K per doubling,
in good agreement with IPCC conclusions based on far more extensive
analysis. The 0.58 K of Ref. \onlinecite{monckton} is very unlikely.
\end{abstract}

 \maketitle

\section{Introduction}
As Spencer Weart has recently noted\cite{weart}, many scientifically trained 
people seek a ``straightforward calculation'' of anthropogenic global 
warming, but ``the nature of the climate system inevitably betrays a lover 
of simple answers.'' The consensus estimates of the climate science
community, expressed by the Intergovernmental Panel on Climate Change (IPCC)
after years of struggling with these complexities, is that 
doubling atmospheric CO$_{2 }$will result in a steady-state temperature 
increase (``equilibrium climate sensitivity'') of 2 to 4.5 K, with a best 
estimate of 3 K, and very unlikely to be less than 1.5 K\cite{IPCCAR4WG1}.

In a recent article in \textit{Physics and Society}\cite{monckton},
Christopher Monckton presents a number of arguments that lead him to conclude
that the IPCC estimate is wrong, and the correct value for
sensitivity should be close to 0.58 K, well within the ``very unlikely"
range stated by the IPCC. I have listed elsewhere\cite{SmithMoncktonErrors}
over 100 errors of fact or logic, misinterpretations, 
invalid reasoning, or misleading statements in Monckton's article.
In many cases these invalid claims are not original with Monckton, but
have been addressed repeatedly in the past. For example he uses at least
ten of the ``hottest skeptic arguments'' found at skepticalscience.com:
``It's the sun'', ``It's cooling'', surface temperatures or models are
unreliable, and so forth. Original with Monckton are a handful of
numerical errors or misinterpretations of others' work; I would refer
those interested in such details to the list I have
collected\cite{SmithMoncktonErrors}; those who spot other errors are 
invited to contact me with details to be added to the list.

The central originality of Monckton's article, which I will address in the 
following, is his breakdown of climate sensitivity into three components, 
and his further reasoning about those components. Monckton's breakdown is 
implicit in much of the discussion about forcings and feedbacks in the IPCC 
report\cite{IPCCAR4WG1} and review papers such as the one by
Bony et al\cite{BonyFeedback}. However, 
it must be emphasized that this breakdown into forcing, base response, and 
feedback factors is artificial and strictly valid only in a perturbative 
sense. It is useful for understanding what is going on, but it is not how 
sensitivity is actually determined. The IPCC consensus on sensitivity comes 
from model calculations and attribution studies that look at the response of 
the full climate system to changes in greenhouse gases, without breaking 
that response into a linearized ``base'' response and separate feedbacks. 
That is, the numerical values for base response and feedbacks, and to a 
lesser degree for forcings, the central concerns of Monckton's article, are 
an output from the numerical models, not an input. They are a guide to 
understanding what is going on, and little more. So while Monckton's 
breakdown is useful in the sense of being more explicit than is typically 
done, it is not in any sense a replication of the manner in which 
sensitivity is determined by the IPCC, since he does not derive his numbers 
from climate models. In fact in his treatment of the 2007 IPCC numbers he 
manages to get both forcing and feedbacks off by 10 to 20{\%}, while coming 
to roughly the right final value\cite{SmithMoncktonErrors}.

\section{Sensitivity Reconsidered, Reconsidered}
But it is the ``reconsidered'' sections of Monckton's article which merit 
the most attention. The entire ``radiative forcing reconsidered'' section 
argues not about forcing at all, but about the temperature changes expected 
from forcings. Monckton ends by claiming he can divide the 3.7 W/m$^{2}$ 
forcing from doubling CO$_{2}$ by a factor of 3 because a certain 
temperature response is low -- but this has nothing to do with the forcing 
at all, which is completely determined by the underlying physics. If 
anything, this section is an argument about feedbacks -- but it is not 
phrased in that way, and so at the least is highly confusing.

Monckton culls a figure from the latest IPCC report (Monckton's figure 4, 
IPCC AR4 WG1 figure 9.1\cite{IPCCAR4WG1}) which is discussed at
length in section 9.2.2.1, ``Spatial and Temporal Patterns of Response'',
but then misinterprets it.
The image is based on estimates of forcing changes from 
1890 to 1999. The strongest pattern is in the greenhouse gas image, because 
that is where the largest forcing change has occured. But the surface and 
low-altitude warming (or cooling) patterns are essentially the same across 
all the forcings -- as the IPCC discussion puts it: ``Solar forcing results 
in a general warming of the atmosphere with a pattern of surface warming 
that is similar to that expected from greenhouse gas warming, but in 
contrast to the response to greenhouse warming, the simulated solar-forced 
warming extends throughout the atmosphere.'' The spatial pattern of response 
between different forcings differs only in the contrast between lower 
atmosphere and upper atmosphere: for solar forcing the warming happens 
everywhere, while for greenhouse forcing the upper atmosphere (stratosphere) 
cools while the surface and lower atmosphere warm. This differential in 
temperature change is readily observed, as Monckton's figure 6 shows: 
warming at the surface, and cooling in the stratosphere. This is 
observational proof that the sun cannot be behind recent warming.

The tropical mid-troposphere ``hot spot'' that Monckton highlights is not a 
``fingerprint'' of greenhouse gases: it is well known to be a consequence of 
higher water vapor levels in a warmer world, whatever the cause of the 
warming. As warm air rises, it cools almost adiabatically -- this is known 
as the ``lapse rate'', and stability of the atmosphere ensures that 
temperatures fall no faster than this rate with altitude. When air holding 
water vapor rises and cools, some of the water condenses and releases heat, 
resulting in warmer air at a given altitude, and a lower lapse rate. The 
strongest effect should show up in the tropical mid-troposphere\cite{SanterTropics}, hence 
a ``hot spot''. The observations are still being disputed, as even Monckton 
admits by refering to the wind-based measurements of Allen et al. Whatever 
the measurements and theory sort themselves out to on this, note again that 
tropical mid-troposphere temperature trends are not a signature of 
greenhouse gases, and this whole argument has no bearing on CO$_{2}$ 
forcing. Monckton has not made any case for arbitrarily dividing the forcing 
by 3 as in his equation 17. That would require drastically changing the 
spectroscopic properties of atmospheric constituents, for which there is 
certainly no justification in the arguments presented.

\begin{figure}[b]
\centerline{\includegraphics[width=6in]{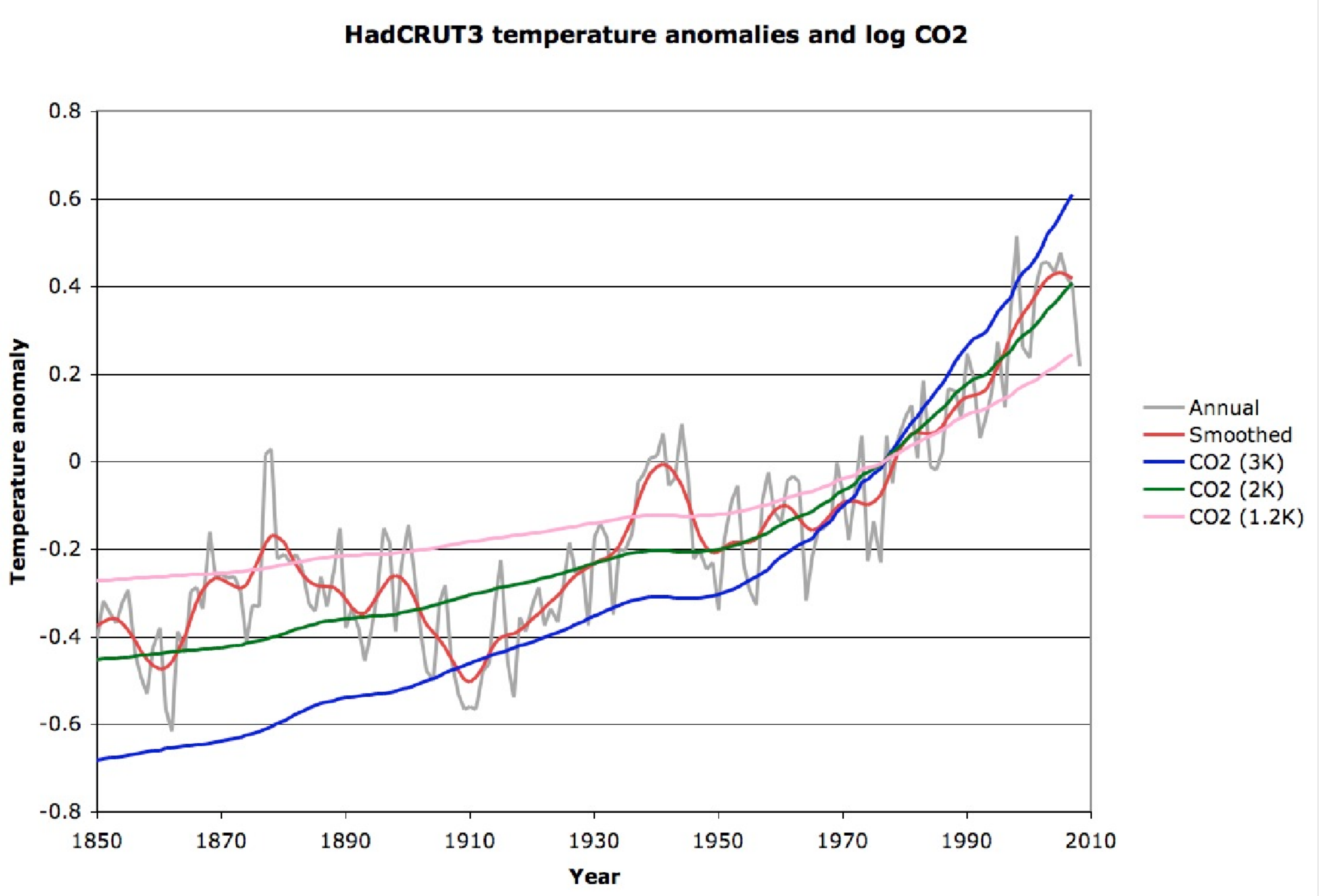}}
\caption{The gray curve is annual average temperature anomaly from the Hadley 
center\cite{HadCrut3}, and in red is the corresponding 21-point smoothed curve. The 
other three lines plot the logarithm of CO$_{2}$ concentrations measured at 
Mauna Loa\cite{MaunaLoa} and the 20-year smoothed Law Dome measurements\cite{LawDome} 
multiplied by appropriate sensitivity-to-doubling numbers and adjusted to a 
0 average for the HadCRUT3 baseline period 1961 to 1990. Other than the 
sensitivity number there is no free parameter here; this is entirely derived 
from observations.}
\label{fig1}
\end{figure}

In his second ``reconsidered'' section, Monckton's errors of logic and 
interpretation are spread thickest, on a matter for which there is again no 
real dispute. Similar to (in fact much more so than for) the forcing, the 
``base'' response of the climate system is tightly constrained by the 
spectroscopic properties and temperature profile of the atmosphere, and is 
easily calculated in any model. The Soden and Held (2006) review paper 
refered to by Bony et al\cite{BonyFeedback} provides two tables which show this base or 
``Planck'' response as calculated from a variety of models. The range of 
what amounts to the inverse of Monckton's $\kappa $ parameter is from 3.13 
to 3.28 W/K m\^{}2, with a mean of 3.22 and standard deviation of 0.04, or 
just over 1{\%}. There is almost no uncertainty over the value of this 
number, despite the confusion Monckton fosters.

Finally, on the feedback factor f, and in particular the sum b of the 
feedback parameters, Monckton argues that the individual feedbacks must be 
too high because adding them plus their standard deviations leads to 
instability. He also quotes two papers that suggest that water vapor and 
cloud feedbacks are overstated by the climate models. The argument of the 
first ``reconsidered'' section on forcings, while completely irrelevant to 
forcings, if it had any validity would reinforce the suggestion of a reduced 
value for feedbacks. Nevertheless, Monckton here does the most mystifying 
thing in his entire article -- he is ``prudent and conservative'' and 
retains the same (excessively high) value for the feedback sum b he has been 
using throughout the article.

But it is the feedbacks that are the most scientifically uncertain issues, 
by far the hardest things for modeling to get right, and the reason you 
cannot determine climate sensitivity from a one-page simple straightforward 
calculation. Getting the complex water vapor, cloud, lapse rate (convection 
and latent heat) and other responses to temperature changes sorted out is 
what makes climate modeling so tough. The feedbacks are the source of almost 
the entire uncertainty range in the IPCC's estimate of climate sensitivity 
(which also relies on measurements of recent and ancient climate). Somehow 
Monckton treats the least certain quantity of the three in his breakdown as 
the most certain, while wildly reducing the values of the other two. His 
final estimate (equation 30) is not believable on these and other grounds, 
including his many other errors\cite{SmithMoncktonErrors}.

\section{Simpler Evidence}
The simplest evidence I have seen that feedbacks are likely to be positive 
comes not from calculations but from measurements, dependent on the IPCC's 
radiative forcing calculations (table 2.12\cite{IPCCAR4WG1}) where the forcings from 
1750 to 2005 due to all sources except CO$_{2}$ almost cancel out (the net 
total is about 10{\%} of the CO$_{2}$ effect, with considerable 
uncertainty). That means an estimate of transient climate response to 
increased CO$_{2}$ can be found by plotting the historically measured 
atmospheric CO$_{2}$ values on the same chart as the historical temperature 
anomalies, over the past 150 years as is done in Figure \ref{fig1}.

The best fit to observed temperature is for a transient response of about 2K 
per doubling of CO$_{2}$. This compares well with the IPCC range of 
transient climate response of between 1 K and 3.5 K (see section 9.6.2.3 of 
IPCC AR4 WG1 \cite{IPCCAR4WG1}. The fact that the 20$^{th}$ century rise in 
temperatures was of almost exactly the expected size is pretty strong 
evidence that the IPCC's transient and equilibrium climate sensitivity 
numbers match reality. In particular, the equilibrium sensitivity is 
unlikely to be as low as the 1.2 K found with no feedbacks, and nowhere near 
the 0.58 K that Monckton claims.

\section{Acknowledgments}
I am grateful to several friends and colleagues for comments; this work has 
received no funding or support from any source.

\end{document}